\documentclass[%
 reprint,
 amsmath,amssymb,
 aps,
 prb,
]{revtex4-2}

\usepackage{graphicx}
\usepackage{dcolumn}
\usepackage{bm}

\usepackage{amsmath,amssymb,amsfonts,mathrsfs}
\usepackage{dsfont}
\usepackage[amsmath,thmmarks]{ntheorem}
\usepackage{xcolor}

\usepackage{defs}

\begin{document}

\preprint{APS/123-QED}

\title{Adiabatic Bottlenecks in Quantum Annealing and Nonequilibrium Dynamics of Paramagnons}

\author{Tim Bode$^{1,}$}
\email{t.bode@fz-juelich.de}
\author{Frank K. Wilhelm$^{1, 2}$}%
 
\affiliation{%
 $^{1}$Institute for Quantum Computing Analytics (PGI-12), Forschungszentrum Jülich, 52425 Jülich, Germany\\
 $^{2}$Theoretical Physics, Saarland University, 66123 Saarbrücken, Germany\\
}%

\date{\today}

\begin{abstract}
The correspondence between long-range interacting quantum spin glasses and combinatorial optimization problems underpins the physical motivation for adiabatic quantum computing. On one hand, in disordered (quantum) spin systems, the focus is on exact methods such as the replica trick that allow the calculation of system quantities in the limit of infinite system and ensemble size. On the other hand, when solving a given instance of an optimization problem, disorder-averaged quantities are of no relevance, as one is solely interested in instance-specific, finite-size properties, in particular the optimal solution. Here, we apply the nonequilibrium Green-function formalism to the spin coherent-state path integral to obtain the statistical fluctuations and the collective-excitation spectrum along the annealing path. For the example of the quantum Sherrington-Kirkpatrick spin glass, by comparing to extensive numerically exact results, we show that this method provides access to the instance-specific bottlenecks of the annealing protocol.
\end{abstract}

\maketitle

\section{Introduction}
\label{sec:intro}

Annealing is the process of very carefully cooling down a physical system. It is an everyday observation that, to make particularly clear, glass-like ice cubes, one has to perform the freezing especially slowly. Fast freezing will result in clouded ice. Quantum annealing, in turn, is the analogous process performed by removing transverse magnetic fields in disordered quantum spin systems, a paradigmatic model of which is the quantum Sherrington-Kirkpatrick (SK) model~\cite{mezard1987spin}. In such systems, the presence of opposing interactions leads to frustration that prevents the ground state from developing simple (anti-) ferromagnetic order. Instead, the system will settle into a ``glass phase''~\cite{mezard2009information} characterized by the existence of many low-energy states masking the true ground state~\cite{mackenzie1982}. Apart from the immediate scientific interest in understanding the properties of spin glasses, their relevance stems from the fact that many combinatorial optimization problems can be cast into a similar form~\cite{mezard2009information}. The complexity of the low-energy landscape then translates into the difficulty of solving the equivalent optimization problem. As the transverse magnetic field is slowly annealed, the most difficult problems are those undergoing a first-order quantum phase transition (QPT)~\cite{choi2009, Young_2010} as they cross over from the delocalized paramagnetic to the localized spin-glass phase. At the critical point of the QPT, the minimal gap between the instantaneous ground state and the first excited state is the crucial quantity determining the precise meaning of \textit{slow} via the adiabatic theorem~\cite{amin2009}. 

\begin{figure}[htb!]
	\begin{center}
		\includegraphics[width=\linewidth]{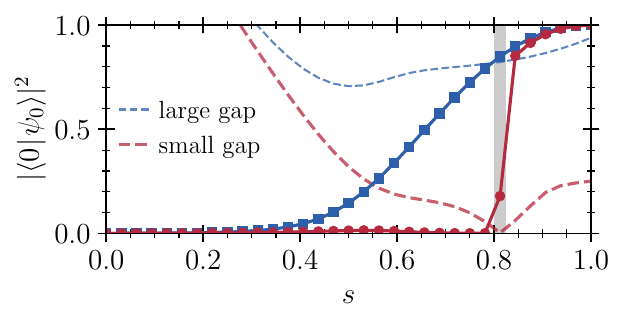}
		\vspace{-0.8cm}\caption
		{\label{fig:fidelity_N_11_seeds_2809_100061} Comparison between the ground-state fidelities of easy and hard realizations of a quantum spin glass. The final ground state of the classical SK model is denoted by $|\psi_0\rangle$, the instantaneous ground state of the adiabatic Hamiltonian in Eq.~\eqref{eq:adiab_ham} by $|0\rangle$. The dashed lines show the instantaneous spectral gaps. The hard instance is defined by the fidelity jump across the minimal gap, which can be interpreted as the finite-size analog of a first-order quantum phase transition~\cite{Werner2023}.} 
	\end{center}
\end{figure}

The recent experimental realizations of quantum-critical spin-glass dynamics~\cite{Amin2023, king2024computational} can be considered a promising step forward, yet it also shows that current experimental values of transverse-field strengths and quantum-coherent final times for superconducting quantum-annealing devices are on the order of $10\,\mathrm{GHz}$ and $ 100\,\mathrm{ns}$, respectively, i.e.\ one can expect final annealing times to be around $10^3$ in units of the initial transverse field. For hard instances of combinatorial optimization problems, the timescale set by the inverse minigap near the critical point will typically be much larger.

This leads us back to the ``most fundamental problem'' of quantum annealing~\cite{choi2009}: how to obtain some information about the size and, in particular, the \textit{onset} of the minigap \textit{without} solving the problem. Here, we demonstrate that the semi-classical fluctuations around the spin mean-field introduced previously~\cite{Misra_2023} can serve as an indicator of the critical point. Physically, these fluctuations describe the dynamics of paramagnons away from equilibrium. They are obtained by solving the equations of motion~\eqref{eq:G_g_l} of the nonequilibrium Green functions~\cite{Altland_2010, kamenev2023field}, which in turn are derived from the Gaussian approximation of the spin coherent-state path integral~\cite{Stone_2000, Altland_2010, Misra_2023}. 

Our main result is the observation that the paramagnons of the most frustrated spins grow systematically around the critical point of the quantum annealing dynamics as defined via the adiabatic theorem~(Figs.~\ref{fig:metric_N_11_seed_2809} and~\ref{fig:disorder_stats}).

To obtain numerically exact data as a comparison for our semi-classical methods, we perform very extensive simulations. Specifically, we diagonalize a large number of random SK instances at different system sizes and retain only the subset with the smallest minimal spectral gaps. The instances in this subset then all contain bottlenecks of varying criticality when annealing the transverse field. The details of these data are provided in Appendix~\ref{sec:app_SK}. Interestingly, even within this critical subset, there remain instances for which the adiabatic mean-field evolution actually finds the true ground state. Our main analysis, however, is then further restricted to those critical instances for which the mean-field evolution \textit{fails}. This appears natural since we want to investigate problems hard for quantum annealing -- but any instance solved via mean-field can be considered easy. Importantly, we find that the information provided by our semi-classical approximation about the critical point of the annealing schedule is \textit{robust}: the method works even when the mean-field algorithm fails, such that we succeed in showing that the alignment between the critical fluctuations and the adiabatic bottleneck is preserved also on average (Fig.~\ref{fig:disorder_stats}).

It has been shown that, on one hand, there exist examples of first-order QPTs with algebraically small minigaps, while on the other hand, one can find rather simple models with exponentially small minigaps~\cite{laumann2012, Hen_2011}. As argued very recently~\cite{bernaschi2023quantum}, it is possible to access the quantum spin-glass phase with an annealing time that only grows as a power law in the number of qubits if parity-changing excitations can be avoided; a possible caveat to this hopeful perspective is that crossing the critical point is not the same as passing through the bulk glass phase up to the final point at zero transverse field, which may still prove difficult. For the quantum SK model in particular, it has been shown that its ``deep glass phase'' is characterized by an Ohmic spectrum of collective excitations~\cite{Mueller_2012}, i.e.\ a gapless phase in the regime of vanishing transverse field. A complementary perspective on this glass phase is provided in Ref.~\cite{Mukherjee_2018}, which demonstrates the existence of a many-body delocalization-to-localization transition as the transverse field is removed. We also note that, based on the idea of tricriticality of the spin-glass phase and localized and delocalized paramagnetic phases, an iterative algorithm has been proposed~\cite{Wang2022} that extends the standard annealing paradigm.

This work is organized as follows. In section~\ref{sec:methods}, we first introduce the quantum version of the SK model as a paradigmatic example of a quantum spin glass, followed by the mean-field approximation (\ref{subsec:mean-field}) and the spin-coherent states that allow us to define the Gaussian fluctuations around this mean-field (\ref{subsec:flucs}). Next, we give a brief overview of the Schwinger-Keldysh formalism (\ref{subsec:schwinger_keldysh}) and nonequilibrium Green functions (\ref{subsec:NEGF}), which are our tools of choice to actually compute the semi-classical dynamics in full generality. The technical part of this paper is then concluded by a discussion of the role of frustration in our quantum spin glasses (\ref{subsec:max_frust}). The results are grouped into two sections, the first of which (\ref{subsec:illustration}) focuses on illustrating the physical ideas leading up to our main result, which is presented in section~\ref{subsec:stats}. A discussion of these results (\ref{sec:conclusion}) concludes this paper.

\begin{figure}[htb!]
	\begin{center}
		\includegraphics[width=0.95\linewidth]{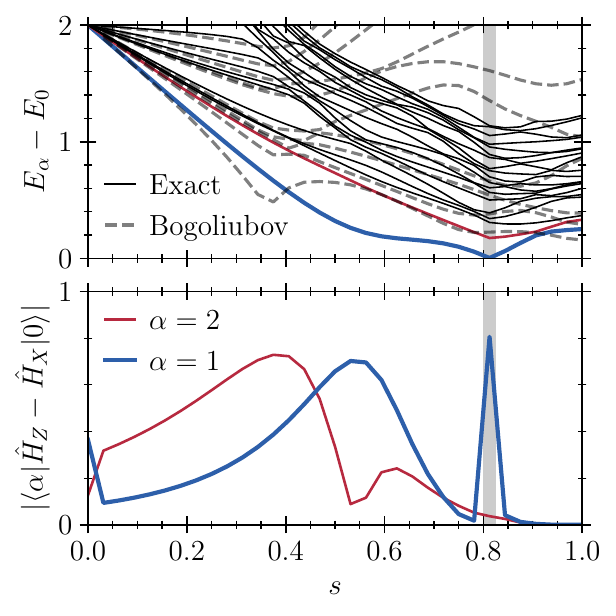}
		\vspace{-0.6cm}\caption
		{\label{fig:exact_spectrum_N_11_seed_2809} The exact eigenvalue spectrum (upper panel) and the couplings between the first two excited states and the ground state (lower panel) as functions of time for a typical hard instance at $N=10$. To obtain the Bogoliubov spectrum (dashed lines in upper panel), Eqs.~\eqref{eq:mf_eom} were solved up to a final time $T_f = 2^{15}$. The exact and Bogoliubov spectra are generated with an economical resolution of $1/32$, which results in $s_*\approx 26/32 = 0.8125$ with $E_1(s_*) - E_0(s_*) = 3.5\cdot 10^{-3}$. The final ground-state energy is $E_0(T_f) = -6.065$ while mean-field converges to the \textit{third} excited state with $E_3(T_f) = - 5.698$.} 
	\end{center}
\end{figure}

\section{Methods}
\label{sec:methods}

We consider a classical spin glass $\hat H_Z$ immersed in a transverse field $\hat H_X$, where ($\hbar=1$)
\begin{align}\label{eq:H_Pand_H_D}
\hat H_Z = - \sum_{i=1}^N \bigg[ h_i  + \sum_{j>i} J_{ij}  \hat\sigma^z_j \bigg] \hat\sigma^z_i, \quad \hat H_X = - \Gamma\sum_{i=1}^N \hat\sigma^x_i.
\end{align}
The couplings $J_{ij}$ and the local magnetic fields $h_i$ are i.i.d.\ standard normal random variables (s.~Appendix \ref{sec:app_SK}). We set $\Gamma = 1$ everywhere, i.e.\ we consider frequencies to be given in units of $\Gamma$. The adiabatic time is confined to $ 0 \leq t \leq T_f$, which makes it natural to also introduce the scaled time $s(t) = t/T_f$, $s(t) \in [0, 1]$. This results in the adiabatic Hamiltonian
\begin{align}\label{eq:adiab_ham}
    \hat H(s) = \rbs{1 - s(t)}\hat H_X + s(t) \hat H_Z.
\end{align}
The time-dependent gap between the instantaneous ground and first excited states will be denoted by $\Delta(s) = E_1(s) - E_0(s)$, the minimal value of which gives the \textit{minigap} $\Delta := \Delta(s_{\mathrm{min}})$, where $s_{\mathrm{min}}$ is the location of the minigap along the annealing path. We also introduce the related quantity
\begin{align}\label{eq:adiabatic_ratio}
    \begin{split}
        g_\alpha(s) = \frac{|\langle \alpha|\partial_s \hat H(s)|0\rangle|}{\sbs{E_\alpha(s) - E_0(s)}^2} = \frac{|\langle \alpha| \hat H_Z - \hat H_X|0\rangle|}{\sbs{E_\alpha(s) - E_0(s)}^2},
    \end{split}
\end{align}
which we will refer to as the \textit{adiabatic ratio}. An illustration of this can be obtained from Fig.~\ref{fig:exact_spectrum_N_11_seed_2809}, where the numerator of Eq.~\eqref{eq:adiabatic_ratio} is plotted in the lower panel while the denominator is given by the inverse square of the upper panel, as well as from Fig.~\ref{fig:metric_N_11_seed_2809}, where $g_1(s)$ is shown directly. The critical point of the annealing dynamics will then be given by
\begin{align}\label{eq:crit_point}
    s_* := \underset{s\in [0, 1]}{\arg\,\max}\;g_\alpha(s),
\end{align}
where typically $\alpha=1$ for the hardest instances. These are almost always very similar to our example of a first-order crossover highlighted in Figs.~\ref{fig:fidelity_N_11_seeds_2809_100061} and \ref{fig:exact_spectrum_N_11_seed_2809}; in particular, the spike in the numerator of $g_1(s)$ plotted the lower panel of Fig.~\ref{fig:exact_spectrum_N_11_seed_2809} coincides characteristically with the minimal gap in the upper panel. This defines the bottleneck according to the adiabatic theorem~\cite{amin2009}, which demands that
\begin{align}\label{eq:adiab_thm}
   \dot{s}(t) g_\alpha(s_*) = \frac{g_\alpha(s_*)}{T_f} \ll 1.
\end{align}
Bottlenecks induced by the \textit{second} excited state $\alpha=2$ can be observed if and when the numerator of $g_1(s)$ happens to vanish around the location of the minigap.

\subsection{Mean-Field Dynamics}
\label{subsec:mean-field}

\begin{figure}[htb!]
	\begin{center}
		\includegraphics[width=\linewidth]{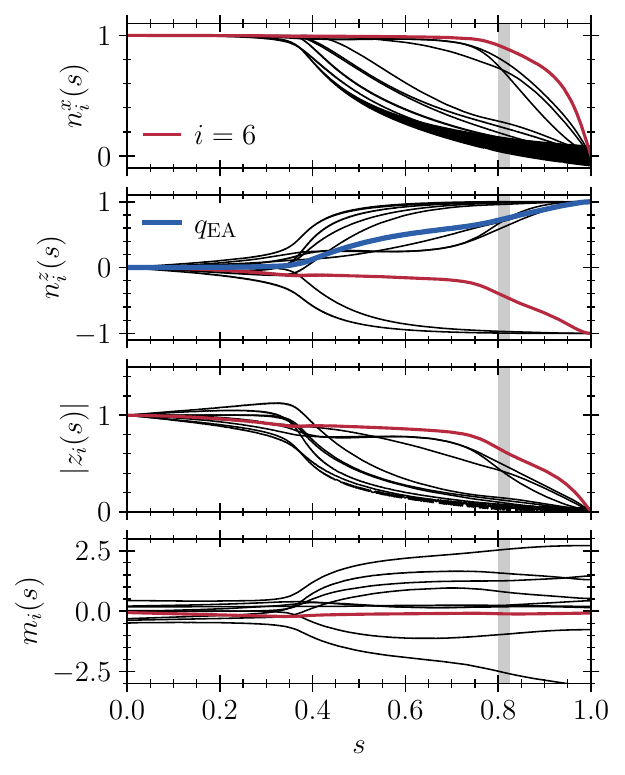}
		\vspace{-0.7cm}\caption
		{\label{fig:mean_fields_N_11_seed_2809} From top to bottom: the $x$- and $z$-components of the mean-fields, the corresponding complex coordinates $z_i$, and the local magnetization $m_i$. Note that we do not show the $y$-components because of Eq.~\eqref{eq:small_n_y}. The Edwards-Anderson order parameter $q_{\mathrm{EA}}$ defined in Eq.~\eqref{eq:EA_param} is also plotted. The most critically frustrated spin $i_* = 6$, which has $n^x_{i_*} \approx 1$ (and accordingly $n^z_{i_*} \approx 0$) right until the occurrence of the critical point around $s\approx 0.8$, is clearly discernible. The final time is $T_f = 2^{15}$.} 
	\end{center}
\end{figure}

The derivation of the mean-field approximation has been detailed in Ref.~\cite{Misra_2023}. The mean-field form of the adiabatic Hamiltonian~\eqref{eq:adiab_ham} is given by
\begin{align}
    \begin{split}
        H(t) ={} &- s(t)\sum_{i=1}^N \bigg[ h_i + \sum_{j>i} J_{ij}  n^z_j(t) \bigg] n^z_i(t) \\
        &- \left(1 - s(t)\right) \sum_{i=1}^N  n^x_i(t),
    \end{split}
\end{align}
with each of the classical spin vectors living on its own Bloch sphere. The classical equations of motion for the spin vectors can be derived in several ways, resulting in
\begin{align}\begin{split}\label{eq:eom_mf}
    \partial_t \boldsymbol{n}_i(t) = \boldsymbol{n}_i(t) \times \boldsymbol{B}_i(t),
\end{split}\end{align}
where $\boldsymbol{B}_i(t) = 2(1 - s(t)) \boldsymbol{\hat{e}}_x + 2s(t)m_i(t) \boldsymbol{\hat{e}}_z$, and the effective local magnetic field is defined as
\begin{align}\begin{split}\label{eq:magnetization}
m_i(t) = h_i + \sum_{j=1}^N J_{ij} n^z_j(t).
\end{split}\end{align}
Written out explicitly in terms of components, these equations become
\begin{align}\begin{split}\label{eq:mf_eom}
\dot n^x_i(t) &= \phantom{-}2 s(t) m_i(t) n^y_i(t),\\
\dot n^y_i(t) &= -2 s(t) m_i(t) n^x_i(t) + 2 (1 - s(t))  n^z_i(t) , \\
\dot n^z_i(t) &= -2 (1 - s(t))  n^y_i(t),
\end{split}\end{align}
i.e.\ we have a set of $3N$ non-linear ordinary differential equations. The norm of all spin vectors is conserved under this evolution, $\left|\boldsymbol{n}_i(t)\right|^2 = 1$. A typical solution of Eqs.~\eqref{eq:mf_eom} for a hard SK instance with $N=10$ is shown in Fig.~\ref{fig:mean_fields_N_11_seed_2809}. 

For very slow evolution, one finds that the trajectories following from Eqs.~\eqref{eq:mf_eom} have the property
\begin{align}\label{eq:small_n_y}
    (n^y_i(t))^2 \ll (n^x_i(t))^2 + (n^z_i(t))^2,
\end{align}
i.e.\ the Bloch vectors are (almost) confined to the $x$-$z$ plane (the real axis of the stereographic complex plane introduced in the next section). Naturally, the trajectories may make large excursions away from the real axis that average out approximately. In fact, this should be the case around critical points (or more generally wherever fluctuations are large). We also note that there is a typical separation of timescales in the dynamics of Eqs.\eqref{eq:mf_eom}, at least when applied to the SK model. The $x$- and $z$-components perform a slow evolution as the vectors move from the equator to either of the poles. Superimposed on this are very fast dynamics of negligible amplitude. The $y$-components, in turn, usually only show the latter. 

We solve Eqs.~\eqref{eq:mf_eom} numerically via the Tsitouras algorithm~\cite{TSITOURAS2011770} implemented in \texttt{DifferentialEquations.jl}~\cite{rackauckas2017differentialequations, bezanson2017julia}. It is advantageous to use such an \textit{adaptive} time-stepping scheme to limit the resources needed to simulate the considerable final times required by adiabaticity. Note, however, that while the mean-field final times should not be too small to ensure convergence, the mean-field dynamics typically converges for much shorter times than those required by Eq.~\eqref{eq:adiab_thm}. This can be understood by considering the Bogoliubov or paramagnon spectrum indicated with the dashed lines in the upper panel of Fig~\ref{fig:exact_spectrum_N_11_seed_2809}: the minimal gap of the instantaneous collective-excitation spectrum is much bigger than the exact one. Accordingly, throughout this work, we typically use final times on the order of $T_f = 2^{15}$ (in inverse units of the transverse field $\Gamma$) with numerical tolerances between $10^{-6}$ and $10^{-8}$. With an eye to our remark about the separation of timescales above, if the only goal is to find a solution bitstring, an efficient way to solve the adiabatic dynamics is to use less strict tolerances, which will usually work well because the low precision will only affect the small-amplitude fast dynamics. In the other direction, the evaluation of the Gaussian fluctuations, introduced below in section~\ref{subsec:flucs}, along the mean-field trajectories can require greater precision when crossing the critical point. 

\subsection{Maximally Frustrated Spins}
\label{subsec:max_frust}

\begin{figure}[htb!]
	\begin{center}
		\includegraphics[width=0.5\linewidth]{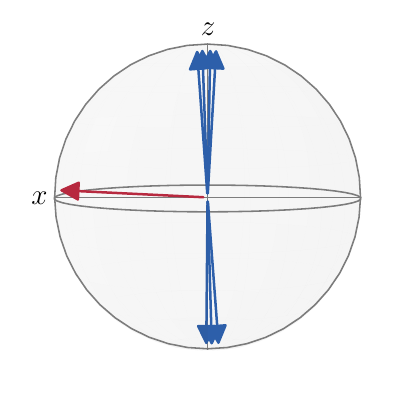}
		\vspace{-0.5cm}\caption
		{\label{fig:bloch_sphere} A graphical illustration of the concept of a highly frustrated spin. When looking at the example trajectories provided in Fig.~\ref{fig:mean_fields_N_11_seed_2809}, we see that the most frustrated spin ($i_*=3$) stays in the equatorial plane until the critical point is reached, while the other spins (except another rather frustrated individual) start converging to $n^z_i = \pm 1$ much earlier. The effective magnetic field $m_{i_*}$ felt by that spin stays close to zero during the entire evolution.} 
	\end{center}
\end{figure}

As described vividly by M{\'e}zard, Parisi, and Virasoro in the introduction to Ref.~\cite{mezard1987spin}, \textit{frustration} is an important and ubiquitous phenomenon. Also in the present context, the tug of opposing forces can have dramatic consequences, leading to strongly frustrated spins that cannot ``decide'' whether to point to the north or south pole of their Bloch spheres. We illustrate this in Fig.~\ref{fig:bloch_sphere}, where one spin fluctuates between the hemispheres around the equator. For our random data set detailed in Appendix~\ref{sec:app_SK}, this can be observed to be quite characteristic of hard instances going through a critical point or the finite-size analog of a first-order phase transition, the onset of which finally forces the frustrated spin either north or south. 

To capture this phenomenon quantitatively, we introduce the local order parameter
\begin{align}\label{eq:local_EA_param}
    \begin{split}
        q_i(s) := n^z_i(s)^2,
    \end{split}
\end{align}
from which the classical Edwards-Anderson (EA) order parameter~\cite{mezard1987spin} follows by summation as
\begin{align}\label{eq:EA_param}
    q_{\mathrm{EA}}(s) = \frac{1}{N}\sum_{i=1}^N q_i(s), 
\end{align}
an example of which is plotted in Fig.~\ref{fig:mean_fields_N_11_seed_2809}. The EA order parameter can be thought of as indicating the degree of localization of the system. Analogously, the amount of \textit{frustration} of a given spin can be determined via the cumulative local order parameter
\begin{align}\label{eq:cum_EA_param}
    Q(i) = \int_0^1\dd s\, q_i(s).
\end{align}
We remark that, alternatively, it is also possible to integrate the effective magnetic field, which yields comparable results; yet another option is to track how the ground-state energy $E_0(T_f)$ at final time changes upon flipping each spin -- the most frustrated spin will typically have the lowest change in energy. Here, we will use Eq.~\eqref{eq:cum_EA_param} to define the \textit{maximally frustrated} spin as
\begin{align}
    i_* := \underset{i\in \{1, ..., N\}}{\arg\,\min}\; Q(i).
\end{align}
As we demonstrate below, focusing on this spin allows us to approximately infer the onset of the critical region of the annealing protocol with much higher precision as compared to aggregate quantities involving all spins.

\subsection{Gaussian Fluctuations}
\label{subsec:flucs}

The Gaussian action functional for the semi-classical fluctuations around the mean-field trajectories can be defined as
\begin{align}
\label{eq:S_fluctuations}
\mathcal{S}[{\eta}, {\bar\eta}] = \frac 1 2 \int_0^T \mathrm{d}t\,  \begin{pmatrix} \bar\eta & \eta \end{pmatrix}
\begin{pmatrix}
i\partial_t - A & B \\
B^\dagger & -i\partial_t - \bar A
\end{pmatrix}
\begin{pmatrix} {\eta} \\ {\bar\eta} \end{pmatrix}.
\end{align}
More details on this derivation can be found in Appendix~\ref{app:gauss} and in Ref.~\cite{Misra_2023}. Note that the expression for the semi-classical inverse propagator of the spin mean-field was first given in Ref.~\cite{Stone_2000}. In terms of Cartesian coordinates on the Bloch sphere, the components of the inverse propagator in Eq.~\eqref{eq:S_fluctuations} are given by $B_{ii}(t) = 0$ and
\begin{align}
\begin{split}
\label{eq:AB_diag}
A_{ii}(t) =  \frac{2 (1-s(t)) n^x_i(t)}{1 + (\boldsymbol{\sigma_*})_i^{} n^z_i(t)} + 2 s(t) (\boldsymbol{\sigma_*})_i^{} m_i(t),
\end{split}
\end{align}
where $\boldsymbol{\sigma}_* = \left(\mathrm{sign}(n_1^z(T_f), ..., \mathrm{sign}(n_N^z(T_f) \right)^T$ is the `solution' bitstring obtained from the mean-field dynamics and determines the Bloch-sphere pole from which the stereographic projection of each spin is performed. For the off-diagonal elements, one obtains
\begin{align}\begin{split}
\label{eq:AB_off-diag}
A_{ij}(t) &= -s(t)\, J_{ij} n^+_i(t) n^-_j(t),\\
B_{ij}(t) &= \phantom{-}s(t)\, J_{ij} n^+_i(t) n^+_j(t),
\end{split}\end{align}
where $n_i^{\pm}(t)= (\boldsymbol{\sigma_*})_i^{} n^x_i(t) \pm i n^y_i(t)$, which shows that ${A} = A^\dagger$ is Hermitian and $B =  B^T$ is symmetric.

\begin{figure}[htb!]
	\begin{center}
		\includegraphics[width=0.85\linewidth]{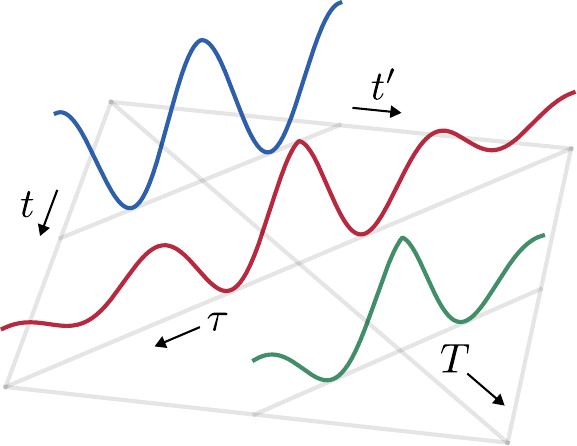}
		\caption
		{\label{fig:3d_elev_35_azim_10} Illustration of the two-time structure of the spectral function $\boldsymbol{\rho}(t, t')$. While the times $t$ and $t'$ are defined along the borders of the temporal square shown in gray, the so-called \textit{Wigner} coordinates $T = \rbs{t + t'}/2$ and $\tau = t - t'$ are defined on the diagonals of the square. The statistical dynamics occurs along the \textit{forward} time $T$, while the Fourier transform of the \textit{relative} time $\tau$ encodes the spectral properties. A corresponding example of $\boldsymbol{\rho}(T, \omega)$ is shown in Fig.~\ref{fig:statistical_and_spectral_N_11_seed_2809}.} 
	\end{center}
\end{figure}

The Gaussian fluctuations are bosonic and thus implicitly define creation and annihilation operators $\hat\eta_i^{{\dagger}}$, $\hat\eta_i^{\phantom{\dagger}}$ for quasi-particles one may imagine as paramagnons. In terms of these operators, we can introduce the so-called \textit{greater} and \textit{lesser} Green functions~\cite{Rammer_2007, stefanucci2013nonequilibrium}
\begin{subequations}\label{eq:greater_lesser_ops}
    \begin{align}
        G_{ij}^>(t, t') &= -\ii\langle \hat\eta_i^{\phantom{\dagger}}(t) \hat\eta_j^{{\dagger}}(t') \rangle, \\
        G_{ij}^<(t, t') &= -\ii\langle \hat\eta_j^{{\dagger}}(t') \hat\eta_i^{\phantom{\dagger}}(t) \rangle.
    \end{align}
\end{subequations}
The diagonals of these two matrices have an intuitive meaning at equal times, i.e.\ for $t' \to t$ (s.\ also Fig.~\ref{fig:3d_elev_35_azim_10}). By letting $\ii G_{ii}^>(t, t) = \mathcal{N}_i + 1$ and $\ii G_{ii}^<(t, t) = \mathcal{N}_i$, we can interpret $\mathcal{N}_i$ as the number of paramagnons (for a bosonic light field, this would just be the number of photons). The only difference between the on-site greater and lesser Green functions at equal times is then given by the zero-point contribution to $G_{ii}^>(t, t)$. By inspecting Eq.~\eqref{eq:S_fluctuations}, we see that the corresponding \textit{anomalous} Green functions proportional to $\langle \hat\eta_i^{\phantom{\dagger}}(t) \hat\eta_j^{\phantom{\dagger}}(t') \rangle$ will also appear in the description of the problem. For completeness, note that the time-dependent Bogoliubov Hamiltonian corresponding to Eq.~\eqref{eq:S_fluctuations} can be written as
\begin{align}
    \begin{split}
        \hat H(t) = \frac 12 \sum_{i,j=1}^N\Big[ &A_{ij}(t)\hat\eta_i^\dagger \hat\eta_j^{\phantom{\dagger}} + A_{ji}(t)\hat\eta_j^\dagger \hat\eta_i^{\phantom{\dagger}} \\[-2mm]
        + &B_{ij}(t)\hat\eta_i^\dagger \hat\eta_j^\dagger + \bar{B}_{ji}(t) \hat\eta_i^{\phantom{\dagger}}\hat\eta_j^{\phantom{\dagger}} \Big].
    \end{split}
\end{align}

\subsection{Schwinger-Keldysh Formalism}
\label{subsec:schwinger_keldysh}

The previous discussions of the fluctuations~\cite{Stone_2000, Misra_2023} have been limited to the corresponding \textit{scattering} problem, where one is interested in computing the transition amplitude between an initial and a final state. The latter introduces a second boundary condition, which has to be handled carefully~\cite{Stone_2000}. Here, we generalize the formalism to the Schwinger-Keldysh contour~\cite{kamenev2023field, Altland_2010, Diehl_Keldysh}, which removes the final condition and allows for a description of the full nonequilibrium dynamics of the semi-classical system along the annealing path.

The nonequilibrium partition function is formally obtained by the well-known `doubling' of degrees of freedom, i.e.\
\begin{align}\label{eq:schwinger_keldysh_partition}
    \mathcal{Z} \sim \int\mathcal{D}\eta^+ \mathcal{D}\bar\eta^+ \mathcal{D}\eta^- \mathcal{D}\bar\eta^-  \exp{\left\{\ii \mathcal{S}[{\eta^+}, {\bar\eta^+}] - \ii \mathcal{S}[{\eta^-}, {\bar\eta^-}]\right\}},
\end{align}
where the minus sign in the exponent originates from the reversal of time on the `backwards' branch of the Schwinger-Keldysh contour. In terms of the contour spinor fields in Eq.~\eqref{eq:schwinger_keldysh_partition}, we may then compactly define all relevant Green functions as
\begin{subequations}\label{eq:greater_lesser_fields}
    \begin{align}
        \boldsymbol{G}^>(t, t') &= -\ii\left\langle \begin{pmatrix} \eta^-(t) \\ \bar\eta^-(t) \end{pmatrix}\begin{pmatrix} \bar\eta^+(t') \hspace{-2mm} & \eta^+(t') \end{pmatrix} \right\rangle, \\
        \boldsymbol{G}^<(t, t') &= -\ii\left\langle \begin{pmatrix} \eta^+(t) \\ \bar\eta^+(t) \end{pmatrix}\begin{pmatrix} \bar\eta^-(t') \hspace{-2mm} & \eta^-(t') \end{pmatrix} \right\rangle.
    \end{align}
\end{subequations}
Finally, in analogy to the equations given previously~\cite{Misra_2023}, the equations of motion for the greater and lesser Green functions then read
\begin{subequations}\label{eq:G_g_l}
\begin{align}
\label{eq:G_t}
        \left[i\sigma_3\overrightarrow{\partial_t} - \mathcal{H}(t)\right]\boldsymbol{G}^\gtrless(t, t') &= 0, \\
\label{eq:G_tp}
        \boldsymbol{G}^\gtrless(t, t')\left[-i\sigma_3\overleftarrow{\partial_{t'}} - \mathcal{H}(t')\right] &= 0,
\end{align}
\end{subequations}
where
\begin{align}
\label{eq:H_t}
\mathcal{H}(t) = \begin{pmatrix}
 A(t) &  B(t) \\ 
 B^\dagger(t) & \bar{A}(t)
\end{pmatrix}, \quad
\sigma_3 =
\begin{pmatrix}
\mathds{1} & \mathds{0}  \\  & -\mathds{1}
\end{pmatrix}.
\end{align}

\subsection{Statistical and Spectral Functions}
\label{subsec:NEGF}

Instead of working with the Green functions defined by Eqs.~\eqref{eq:greater_lesser_ops} or \eqref{eq:greater_lesser_fields}, respectively, it is conventional to transform to the following equivalent linear combinations,
\begin{align}
    \begin{split}
        \boldsymbol{F}(t, t') &= \sbs{\boldsymbol{G}^>(t, t') + \boldsymbol{G}^<(t, t')} \sigma_3, \\
        \boldsymbol{\rho}(t, t') &= \sbs{\boldsymbol{G}^>(t, t') - \boldsymbol{G}^<(t, t')} \sigma_3.
    \end{split}
\end{align}
We will refer to $\boldsymbol{F}(t, t')$ as the \textit{statistical} function, while $\boldsymbol{\rho}(t, t')$ is the so-called \textit{spectral} function. Similarly to our above discussion of the equal-time functions, it now follows that $\ii F_{ii}(t, t) = 2\mathcal{N}_i + 1$ is directly related to the paramagnon number. The spectral function, in turn, has the property $\boldsymbol{\rho}(t, t) = -\ii \mathds{1}$, the origin of which lies in the bosonic commutation relation. An illustration of the behavior of this function away from the equal-time diagonal is provided in Fig.~\ref{fig:3d_elev_35_azim_10}. The equation of motion of the statistical function can be derived easily by combining Eqs.~\eqref{eq:G_g_l} to yield
\begin{align}\label{eq:stat_GF}
    \ii\partial_T \boldsymbol{F}(T, 0) &= \left[\sigma_3 \mathcal{H}(T), \boldsymbol{F}(T, 0)\right],
\end{align}
with initial condition $\boldsymbol{F}(T, 0) = -\ii \sigma_3$. Note that while Eq.~\eqref{eq:stat_GF} is \textit{superficially} equivalent to Eq.~(59) of Ref.~\cite{Misra_2023}, the mathematical problem is not identical: the latter formally belongs to a boundary-value problem, while Eq.~\eqref{eq:stat_GF} together with the initial condition amounts to a standard $2N\times 2N$ initial-value problem.

Furthermore, the solution of the full nonequilibrium dynamics, as developed here, requires two further equations of motion, one for the statistical and one for the spectral function. The former can be derived from Eq.~\eqref{eq:G_t} by adding the respective equations for $\boldsymbol{G}^>$ and $\boldsymbol{G}^<$; it is not relevant here since we only require the equal-time dynamics $\boldsymbol{F}(T, 0) = \boldsymbol{F}(t, t)$, which, in the \textit{Gaussian} case, decouples from the dynamics away from the forward time diagonal $t=t'$ (s.~Fig.~\ref{fig:3d_elev_35_azim_10}). The remaining equation of motion for the spectral function can also be derived straightforwardly from Eqs.~\eqref{eq:G_g_l} and reads
\begin{align}\label{eq:spec_GF}
    \begin{split}
            \ii\partial_\tau \boldsymbol{\rho}(T, \tau) &= \frac 12 \sigma_3 \mathcal{H}\rbs{T + \frac{\tau}{2}} \boldsymbol{\rho}(T, \tau) \\
    &+  \frac 12 \boldsymbol{\rho}(T, \tau) \sigma_3 \mathcal{H}\rbs{T - \frac{\tau}{2}},
    \end{split}
\end{align}
with initial conditions $\boldsymbol{\rho}(T, 0) = -\ii \sigma_3$ for all $T$. Again, it is only in the Gaussian approximation that this equation decouples from the general dynamics on the full two-time square. 

Looking to Fig.~\ref{fig:3d_elev_35_azim_10}, we see that Eq.~\eqref{eq:stat_GF} describes the evolution along the physical \textit{forward-time} diagonal $T = (t+t')/2$, while Eq.~\eqref{eq:spec_GF} models the spectral dynamics in the orthogonal \textit{relative-time} direction $\tau = t-t'$. Fourier transforming $\boldsymbol{\rho}(T, \tau)$ in the second argument provides the nonequilibrium collective-excitation spectrum as a function of $T$. An example of this is plotted in Fig.~\ref{fig:statistical_and_spectral_N_11_seed_2809}.

\section{Results}
\label{sec:results}

\begin{figure*}[htb!]
	\begin{center}
		\includegraphics[width=\textwidth]{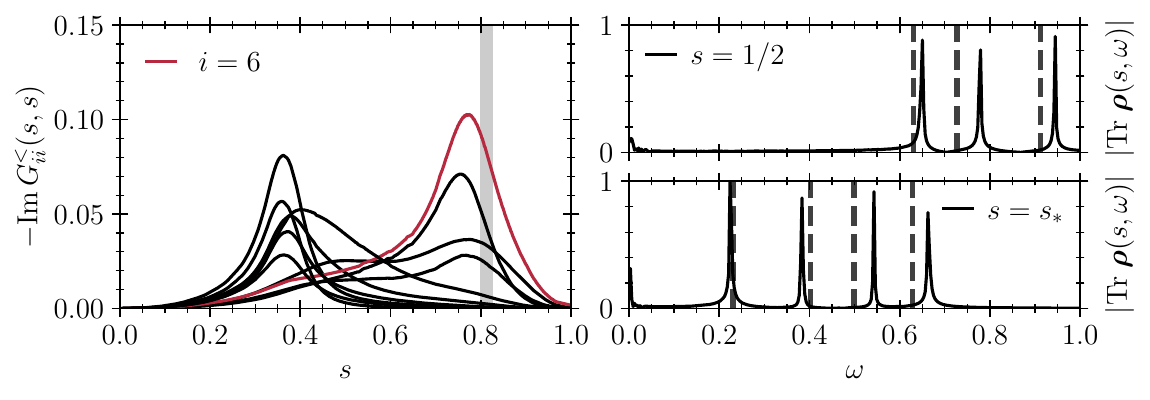}
		\vspace{-0.5cm}\caption		{\label{fig:statistical_and_spectral_N_11_seed_2809} The statistical fluctuations (left) and the collective-excitation spectrum (right) as functions of time. The final relative time for the spectral evolution shown on the right is $\tau_f = 2^{11}$. The forward-time slices are taken at $s = 1/2$ and $s_* = T/T_f = 26/32 = 0.8125$, where $T_f = 2^{15}$ as before. Note that we do not expect perfect agreement between the spectral and Bogoliubov frequencies, which are indicated by the vertical dashed lines.} 
	\end{center}
\end{figure*}

\begin{figure*}[htb!]
	\begin{center}
        \includegraphics[width=\textwidth]{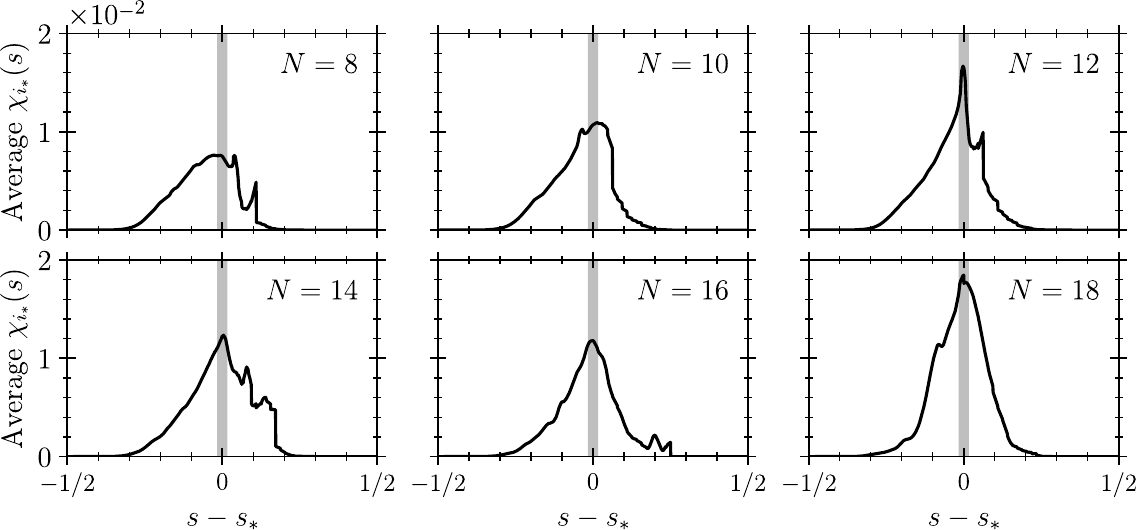}
		\vspace{-0.5cm}\caption{\label{fig:disorder_stats} The disorder-averaged {localization susceptibilities} $\chi_{i_*}(s)$ from Eq.~\eqref{eq:loc_suscep} evaluated for each maximally frustrated spin $i_*$. The critical points $s_*$ are determined from $g_1(s)$ via Eq.~\eqref{eq:crit_point} separately for each instance; then the corresponding localization susceptibilities are shifted. The disorder average is taken \textit{after} this transformation. The effective ensemble sizes for the disorder averages are given in Tab.~\ref{tab:eff_ens}.} 
	\end{center}
\end{figure*}

To illustrate the physical concepts necessary for understanding our main result in Fig.~\ref{fig:disorder_stats}, we will first focus on a single hard SK instance at $N=10$. All the quantities shown in Figs.~\ref{fig:fidelity_N_11_seeds_2809_100061}, \ref{fig:exact_spectrum_N_11_seed_2809} and Figs.~\ref{fig:mean_fields_N_11_seed_2809} to \ref{fig:metric_N_11_seed_2809} are evaluated for this instance, which is chosen from the bulk of the small-minigap instances at $N=10$, i.e.\ it is neither the hardest nor the easiest as measured by the maximum of $g_1(s)$. This illustration is then extended by a statistical analysis of disorder-averaged quantities in section~\ref{subsec:stats}.

\subsection{Illustration: Details of a Hard Instance}\label{subsec:illustration}

\begin{figure}[htb!]
	\begin{center}
		\includegraphics[width=\linewidth]{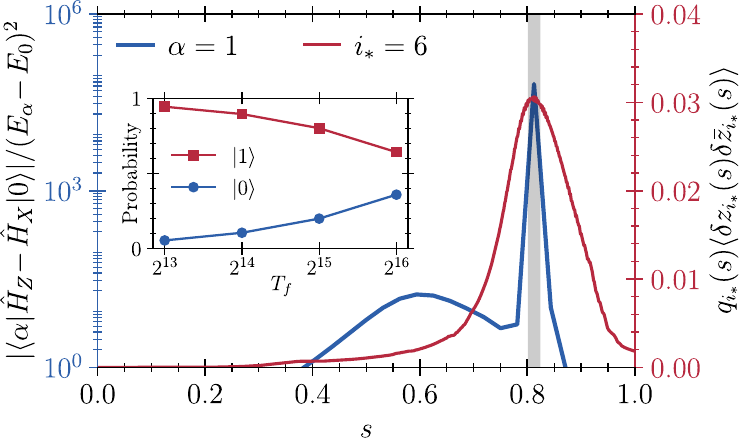}
		\vspace{-0.6cm}\caption
		{\label{fig:metric_N_11_seed_2809} The adiabatic ratio $g_{1}(s)$ (left axis) compared to the localization susceptibility of the maximally frustrated spin $i_* = 6$ (right axis). The contribution from $g_2$ is negligible. The two peaks are almost perfectly aligned with the peak of the adiabatic ratio $g_1$, illustrating that the semi-classical fluctuations are capable of detecting critical points. The inset shows the performance of quantum annealing when simulated via second-order Trotterization~\cite{Willsch2020}.} 
	\end{center}
\end{figure}

The exact ground-state fidelity of the exemplary instance to be considered in this section is shown in Fig.~\ref{fig:fidelity_N_11_seeds_2809_100061}, while the exact spectrum is given in Fig.~\ref{fig:exact_spectrum_N_11_seed_2809}. The critical point is located around $s_*\approx 0.8$, where the spectral gap between the ground and first excited state becomes minimal, which is accompanied by a characteristic peak in the numerator of Eq.~\eqref{eq:adiabatic_ratio}. The corresponding adiabatic ratios $g_{1, 2}(s)$ are obtained by dividing the the curves in the lower by those in the upper panel of Fig.~\ref{fig:exact_spectrum_N_11_seed_2809}. A proper plot of $g_1(s)$ is provided in Fig.~\ref{fig:metric_N_11_seed_2809}, from which we glance an adiabatic bottleneck~\cite{amin2009} on the order of
\begin{align}
    \underset{s\in [0, 1]}{\max}\;g_1(s) \sim 6\cdot 10^4.
\end{align}
With an eye to the current hardware parameters quoted in the introduction~\ref{sec:intro}, obtaining the true solution to this instance should prove a challenge without further counter-diabatic measures~\cite{Lidar_2010, polkovnikov_2017}. We also remark that the mean-field evolution applied to this example instance does \textit{not} find the true solution but instead converges to only the \textit{third} excited state, i.e.\ all conclusions we can draw from the semi-classical approximation derived from these mean-fields are not conditioned on successful adiabatic mean-field dynamics.

The details of the adiabatic mean-field dynamics according to Eqs.~\eqref{eq:mf_eom} are shown in Fig.~\ref{fig:mean_fields_N_11_seed_2809}. The upper two panels show the evolution of the components $n^{x,z}_i(s)$; note that we do not show $n^y_i(s)$ because of Eq.~\eqref{eq:small_n_y}. The complex coordinate $z_i(s)$, which is important in the coordinate transformation defined by Eq.~\eqref{eq:delta_zi}, is shown in the third panel. The last panel illustrates the effective local magnetization Eq.~\eqref{eq:magnetization}. For all of these quantities, the critical spin determined via Eq.~\eqref{eq:cum_EA_param} is highlighted in red. Clearly, it is the spin for which the transition from the delocalized to the localized regime, i.e.\ in the mean-field sense from $n^x_i \approx 1$, $n^z_i \approx 0$ to $n^x_i \approx 0$, $|n^z_i| \approx 1$ happens last and, more importantly, roughly around the point where the ground-state fidelity of Fig.~\ref{fig:fidelity_N_11_seeds_2809_100061} starts to shoot up. By extension, the same will be true for our local order parameter $q_i(s)$ defined in Eq.~\eqref{eq:local_EA_param}. Looking beyond the maximally frustrated spin, we see that the same is true, to a lesser degree, for another group of three spins. Relatedly, the corresponding complex coordinate $z_i$, which provides the scale factor between the Gaussian fluctuations $\eta_i$ and the fluctuations in the stereographic plane $\delta z_i$, is also much larger (relatively) than for the rapidly localizing spins. For this reason, we define the new quantity
\begin{align}
    \begin{split}
        \E{\delta z_i(s) \delta\bar z_i(s)} := \rbs{1 + \left|z_i(s)\right|^2}^2 \frac{\ii}{2} \rbs{F_{ii}(s, s) - 1},
    \end{split}
\end{align}
which one could call the \textit{scaled local fluctuations} and which should provide a more accurate picture of the actual fluctuations of the trajectories on the Bloch sphere. 

For our current example, the total EA order parameter from Eq.~\eqref{eq:EA_param}, plotted in blue in the second panel, does not show a very distinct increase around the critical point, which is due to the fact that the already localized spins have the largest contributions. This underlines the usefulness of focusing on the critical spin. Finally, consider the effective magnetization shown in the last panel of Fig.~\ref{fig:mean_fields_N_11_seed_2809}. Intriguingly, we perceive that our critical spin is completely frustrated during the entire evolution, an observation that we have found repeated throughout the full data set described in Appendix~\ref{sec:app_SK}. For this reason, we refer to the critical spin as \textit{maximally frustrated}.

Having discussed the mean-field quantities, we now move to the semi-classical approximation and consider the results of Eqs.~\eqref{eq:stat_GF} and \eqref{eq:spec_GF}, which are presented in Fig.~\ref{fig:statistical_and_spectral_N_11_seed_2809}. The most important result is that the local statistical fluctuations of the maximally frustrated spin, quantified via $G^<_{ii}$ evaluated at equal times $s = s'$ in the left panel, are largest with a peak quite close to the critical point. Again, this is a typical observation for hard instances and not an accidental feature of our current example. To complete the picture of the nonequilibrium dynamics, the right panels of Fig.~\ref{fig:statistical_and_spectral_N_11_seed_2809} show the spectral function after Fourier transform in its second argument, to which one could also refer as the collective-excitation spectrum. The vertical dashed lines repeat the instantaneous Bogoliubov spectrum from the upper panel of Fig.~\ref{fig:exact_spectrum_N_11_seed_2809}. Note that one should not expect perfect agreement between the latter and the nonequilibrium spectral function, the less so the more the dynamics departs from adiabaticity. It is also important to remark that the line width of this Fourier transform is dictated by the length of the available time intervals in the direction of $\tau = t-t'$ (cf.~Fig.~\ref{fig:3d_elev_35_azim_10}), which is maximal for $t=t'=T_f/2$ but very short both early and late in the annealing protocol. 

Intuitively, the observation that the critical peak in Fig.~\ref{fig:statistical_and_spectral_N_11_seed_2809} is slightly early suggests to consider the quantity
\begin{align}\label{eq:loc_suscep}
    \begin{split}
        \chi_i(s) := q_i(s)\E{\delta z_i(s) \delta\bar z_i(s)},
    \end{split}
\end{align}
to which we will refer as the \textit{localization susceptibility}. It provides a joint measure of how localized and susceptible to fluctuations the maximally frustrated spin is at any given time. The resulting curve for our current example is presented in Fig.~\ref{fig:metric_N_11_seed_2809} alongside the adiabatic ratio $g_1(s)$ as defined in Eq.~\eqref{eq:adiabatic_ratio}. The peaks of the two quantities are almost perfectly aligned, which we consider a promising counterexample to the statement that mean-field dynamics \textit{cannot} address the above-quoted ``fundamental problem'' of quantum annealing. To substantiate this observation, we provide disorder-averaged statistics in the next section. 

To briefly comment on the inset of Fig.~\ref{fig:metric_N_11_seed_2809}, which shows the outcomes of numerically exact simulations of the evolution under Eq.~\eqref{eq:adiab_ham} for various final times $T_f$, we see that the ground state only begins to become visible around $T_f \sim 10^4$, in agreement with the expectation from the adiabatic theorem~\cite{amin2009}.

Regarding the role of the $q_i(s)$ factor in Eq.~\eqref{eq:loc_suscep}, it is helpful to also consider the additional results presented in Appendix~\ref{sec:app_max2sat}, where we apply our method to Maximum 2-Satisfiability (Max.~2-Sat.). While the most frustrated mean-field trajectory of our present example does localize to one of the Bloch-sphere poles at the end of the evolution, this is not always the case, as can be seen from Fig.~\ref{fig:max2sat_mean_fields_N_10}. For this hard Max.~2-Sat.\ instance, while we find that $q_{i_*}(1) = 0$, i.e.\ the most frustrated spin does not localize eventually, our definition of the localization susceptibility remains unaltered and leads to the result shown in Fig.~\ref{fig:max2sat_metric_N_10}.

\subsection{Main Result: Disorder-Averaged Statistics}\label{subsec:stats}

To go beyond the anecdotal evidence presented in the previous section, we now extend the analysis to an extended data set of small-minigap instances, the details of which are set out in Appendix~\ref{sec:app_SK}. The system and ensemble sizes of this data set are summarized in Tab.~\ref{tab:SK_small}. These instances are obtained by computing the numerically exact spectrum and retaining only those that have a minigap $\Delta < \Delta_{\rm small} = 10^{-2}$. As mentioned in the introduction, even among these instances there occur not a few that are solved \textit{exactly} by adiabatic mean-field dynamics. We emphasize that we \textit{exclude} these instances from the following analysis because the whole point is to devise an approximate method applicable to \textit{hard} instances that cannot be solved in mean-field. The resulting effective ensemble sizes are given in Tab.~\ref{tab:eff_ens}. These remaining instances span a range of criticality as quantified by the peak of the corresponding adiabatic ratios.

For each instance in these ensembles, we determine the critical point $s_*$ according to Eq.~\eqref{eq:adiabatic_ratio} and then plot the corresponding maximally frustrated localization susceptibility against $s - s_*$, i.e.\ $\chi_{i_*}(s)$ must peak around the origin in order to have its maximum coincide with the adiabatic bottleneck. Importantly, we perform the disorder average of the random realizations of the SK spin glass \textit{after} this transformation. The results are shown in Fig.~\ref{fig:disorder_stats}: except for the smallest systems at $N=8$, the average localization susceptibility does indeed peak at the origin, which confirms that $\chi_{i_*}(s)$ contains statistically relevant information about the onset of the adiabatic bottleneck. Note that if one only considers the local EA order parameter Eq.~\eqref{eq:local_EA_param}, the averaged agreement with the peak of the adiabatic ratio is worse than when taking the product with the maximally frustrated localization susceptibility. The same is true if considering only the average of the latter. We remark that while the alignment between localization susceptibility and adiabatic bottleneck is not always as ideal as in Fig.~\ref{fig:metric_N_11_seed_2809}, there is almost always good agreement (as is necessary to obtain the significant average agreement observed in Fig.~\ref{fig:disorder_stats}).

\begin{table}[htb!]
\centering
\def\arraystretch{1.25}
\begin{tabular}{c|c|c|c|c|c|c}
$N$ & 8 & 10 & 12 & 14 & 16 & 18 \\
\hline
Ensemble & 2171 & 2169 & 1939 & 1050 & 250 & 110 \\
\hline
Avg.\ Hamming distance & 5.98 & 7.02 & 7.87 & 9.12 & 10.00 & 11.04 \\
\end{tabular}
\caption{Effective ensemble sizes and average Hamming distances to the optimal solution after excluding small-minigap instances solved by adiabatic mean-field evolution.}\label{tab:eff_ens}
\end{table}

\section{Discussion \& Conclusion}
\label{sec:conclusion}

We have developed the theory of the nonequilibrium dynamics of the semi-classical spin coherent-state path integral, as encapsulated by Eqs.~\eqref{eq:mf_eom} and \eqref{eq:G_g_l}. This has allowed us to study the instance-specific onset of the adiabatic bottleneck for the quantum SK model via critical fluctuations. Our main result, Fig.~\ref{fig:disorder_stats}, provides numerical evidence that this is statistically robust and not dependent on the system size. 

It seems clear that there are many instances where the minigap with the first excited state is so small that one cannot hope of mitigating it practically in the near future. However, insofar as a bulk glass phase with a ``soft'' excitation spectrum~\cite{Mueller_2012} can be considered universal for long-range interacting spin glasses, there is the real possibility of doing \textit{even worse} by scattering into higher excited states. This was indeed the case for our example instance above, and it seems to become more likely at larger $N$ (as indicated by Fig.~\ref{fig:mf_sol_hist_all} in Appendix~\ref{sec:app_SK}). A minimal intermediate goal could therefore be to mitigate these higher excitations, a goal for which the knowledge of the critical fluctuations, as presented in this work, appears to be very helpful. 

An interesting direction for future investigation is the question of the connection of our methods to chaos and entanglement growth in quantum many-body systems~\cite{pappalardi2020, lerosePRR2020}. Another question relates to the role of (semi-) classical chaos in the mean-field and fluctuation dynamics: one can imagine a situation where the onset of classical chaos prevents the mean-fields from converging to a properly localized final result. While this would render adiabatic mean-field evolution meaningless as a method to \textit{solve} combinatorial optimization problems, it appears likely that this would also be accompanied by a divergence of critical fluctuations -- which from our perspective would still be useful in the sense of detecting the onset of criticality.

We finally remark that there are immediate implications of our work for annealing-schedule design along the lines of Ref.~\cite{Lidar_2010}. This should be combined with an exploration of the applicability of our method to hard instances of other typical combinatorial optimization problems such as maximum satisfiability~\cite{crosson2014different, Mirkarimi_2023}.

\begin{acknowledgments}
The authors would like to thank Dmitry Bagrets and Krish Ramesh for helpful comments, and acknowledge partial support from the German Federal Ministry of Education and Research, under the funding program “Quantum technologies - from basic research to the market,” Contracts No.\ 13N15688 (DAQC) and No.\ 13N15584 [Q(AI)2].
\end{acknowledgments}

\appendix

\section{Gaussian Fluctuations}\label{app:gauss}

As shown in Refs.~\cite{Stone_2000, Misra_2023}, the Gaussian spin coherent-state path integral for the fluctuations around the mean-field can be understood via stereographic projection, which is defined as the map $\mathcal{P}: \mathds{C}\to S_2$, $z \mapsto (n^x, n^y, n^z)^T  $ from the complex plane to the sphere, where for a single spin one finds
\begin{align}\begin{split}
\label{eq:n_via_z}
n^x \pm \ii n^y = \frac{2 z}{1 + |z|^2}, \quad n^z = \frac{\pm\left(1-|z|^2\right)}{1 + |z|^2}.
\end{split}\end{align} 
The inverse map $S_2 \to\mathds{C}$ reads
\begin{align}\begin{split}
\label{eq:z_via_n}
z = \frac{n^x \pm \ii n^y}{1 \pm n^z}.
\end{split}\end{align}
This projects the north pole $(0, 0, 1)$ (south pole $(0, 0, -1)$) to the origin of $\mathds{C}$ while the south pole (north pole) goes to infinity (the equator becomes the unit circle $|z|=1$). Note that the corresponding path-integral measure is \textit{not flat}, but rather proportional to
\begin{align}
    \int \frac{\mathrm{d}z \mathrm{d}\bar z}{\left(1 + z\bar z\right)^{2}}.
\end{align}
The fluctuations $\delta z_i^{}$ of each spin are now defined in this complex representation, i.e.\ as the deviations of the true trajectories $z'_i$ from the mean-fields $z_i^{}$, such that $z'_i = z_i^{} + \delta z_i^{}$.  Next, by observing the identity
\begin{align}
\label{eq:delta_zi}
\delta z_i = (1+|z_i|^2)\eta_i + \mathcal{O}(\eta_i^2),
\end{align} 
via a change of variables~\cite{Stone_2000} from $\delta z_i$ to $\eta_i$ it is then possible to obtain a Gaussian path integral with a \textit{flat} measure,
\begin{align}
    \mathcal{A} \sim \int\mathcal{D}\eta \mathcal{D}\bar\eta \exp{\left\{\ii \mathcal{S}[{\eta}, {\bar\eta}]\right\}},
\end{align}
where $\mathcal{S}$ is defined in Eq.~\eqref{eq:S_fluctuations}.

\section{Maximum 2-Satisfiability}\label{sec:app_max2sat}

\begin{figure}[htb!]
	\begin{center}
		\includegraphics[width=\linewidth]{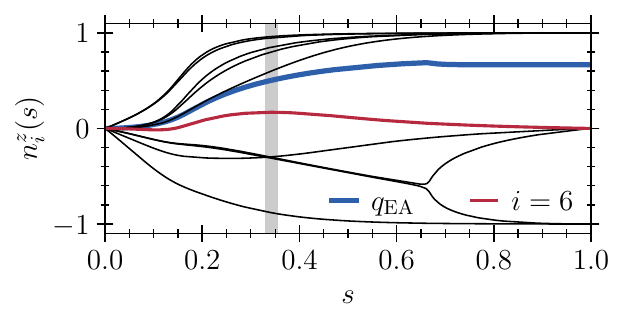}
		\vspace{-0.75cm}\caption
		{\label{fig:max2sat_mean_fields_N_10} Max.~2-Sat.\ mean-field trajectories for $N=10$ and $T_f = 2^{16}$.} 
	\end{center}
\end{figure}

Here we present supplementary results for a hard instance from the maximum-2-satisfiability data set of Ref.~\cite{Mirkarimi_2023}, where the mean-field method converges to an excited state. Notably, the most frustrated spin in Fig.~\ref{fig:max2sat_mean_fields_N_10} does not localize at the end of the evolution.

\begin{figure}[htb!]
	\begin{center}
		\includegraphics[width=\linewidth]{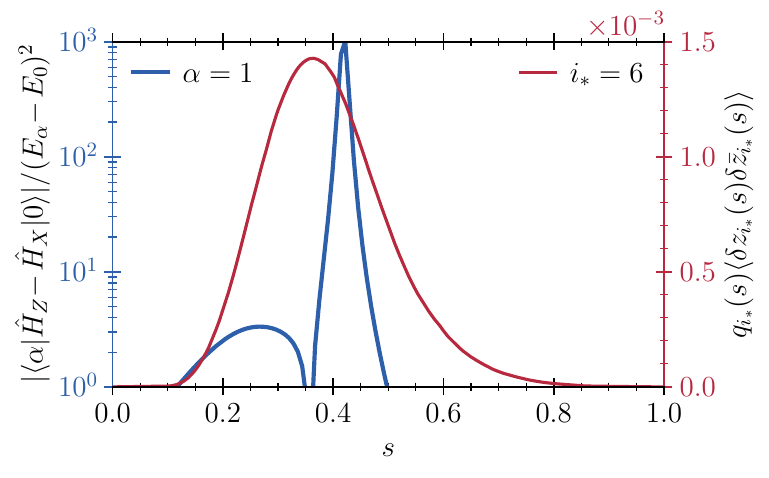}
		\vspace{-0.75cm}\caption
		{\label{fig:max2sat_metric_N_10} Max.~2-Sat.\ localization susceptibility for $N=10$ and $T_f = 2^{16}$.} 
	\end{center}
\end{figure}

This does not render our definition of the localization susceptibility useless, however, as is confirmed by Fig.~\ref{fig:max2sat_metric_N_10}. While we only show one example here to illustrate this point, we have confirmed that these results are robust by considering other instances. 

\section{Sherrington-Kirkpatrick Data Set}\label{sec:app_SK}

Here we give a detailed overview of the data set of random SK instances used in the main text. Tab.~\ref{tab:SK_small} shows the ensemble sizes for the different spin numbers for the instance set with small minimal gaps, where the cut-off for the minigap size was $\Delta_{\rm small} = 10^{-2}$. We generate instances of size $N+1$ with problem Hamiltonians
\begin{align}\begin{split}\label{eq:H_SK}
    \hat H_P = - \frac{1}{\sqrt{N+1}}\sum_{i<j\leq N+1} J_{ij} \hat\sigma^z_i \hat\sigma^z_j,
\end{split}\end{align} 
where the couplings $J_{ij}$ are i.i.d.\ standard normal random variables, i.e.\ with zero mean $\left\langle J_{ij} \right\rangle = 0$ and variance $ \left\langle J_{ij}^2 \right\rangle = 1$. Then we fix $n^z_{N+1} = 1$ to ensure the symmetry breaking necessary for the application of the mean-field approximation~\cite{Misra_2023}. This transforms $\hat H_P$ to $\hat H_Z$ as defined in Eq.~\eqref{eq:H_Pand_H_D}. Note that, however, the \textit{classical} ground state of $\hat H_Z$ is identical to one of the ground states of $\hat H_P$. 

\begin{table}[htb!]
\centering
\def\arraystretch{1.25}
\begin{tabular}{c|c|c|c|c|c|c}
$N$ & 8 & 10 & 12 & 14 & 16 & 18 \\
\hline
Ensemble & 2555 & 2581 & 2307 & 1297 & 310 & 137 \\
\end{tabular}
\caption{The SK data set with small minigaps $\Delta < \Delta_{\rm small} = 10^{-2}$.}\label{tab:SK_small}
\end{table}
From diagonalizing $20 \cdot 10^{3}$ random instances at $N=8$, we obtained 462 small-minigap instances. This can be compared with the roughly $2000$ random instances we had to diagonalize at $N=18$ to find about 140 small-minigap instances. The percentage of hard instances thus increases from about $2.3\%$ at $N=8$ to around $7\%$ at $N=18$, in agreement with the expectations for the SK model. We also retained a set of random instances with large minigaps while searching for these hard instances. The respective ensembles are summarized in Tab.~\ref{tab:SK_large}. As the cost of obtaining the true spectrum obviously becomes prohibitive very quickly as a function of $N$, even the observed rise in hard instances cannot remedy the shrinking in ensemble sizes detailed in Tab.~\ref{tab:SK_small}. Note, however, that we observed a gradual improvement of our main result in Fig.~\ref{fig:disorder_stats} as we slowly increased the ensemble sizes of the larger values of $N$, i.e.\ we have good reason to believe that our findings are robust and would indeed \textit{improve} as more instances are added.

\begin{table}[htb!]
\centering
\def\arraystretch{1.25}
\begin{tabular}{c|c|c|c|c|c|c}
$N$ & 8 & 10 & 12 & 14 & 16 & 18 \\
\hline
Ensemble & 500 & 500 & 500 & 500 & 50 & 20 \\
\end{tabular}
\caption{The SK data set with large minigaps $\Delta > \Delta_{\rm large}(N)$.}\label{tab:SK_large}
\end{table}

\begin{figure}[htb!]
	\begin{center}
		\includegraphics[width=\linewidth]{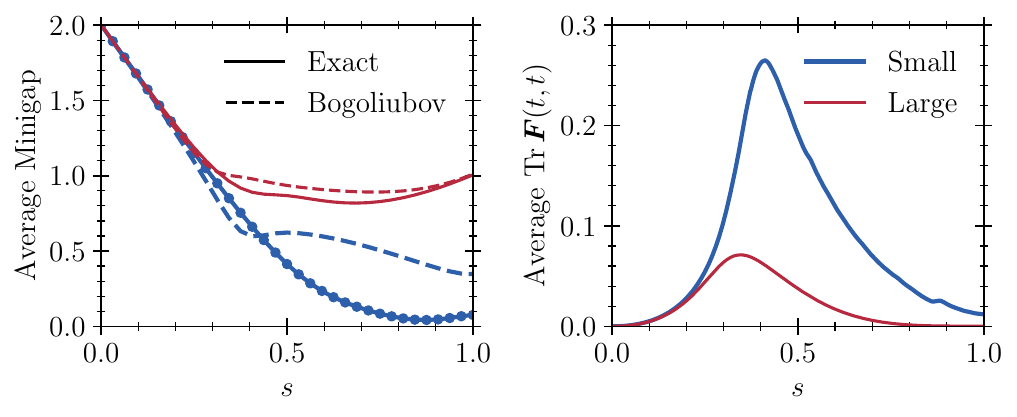}
		\vspace{-0.75cm}\caption
		{\label{fig:gap_flucs_N_9} (Left) Ensemble-averaged minigaps for the two data sets at $N=8$. (Right) Ensemble-averaged sum over the statistical fluctuations for a final time $T_f=32000$ with $\Delta t = 2^{-11}$. Observe that there is a small secondary peak in the fluctuations at the location of the minigap.} 
	\end{center}
\end{figure}

\begin{figure}[htb!]
	\begin{center}
		\includegraphics[width=\linewidth]{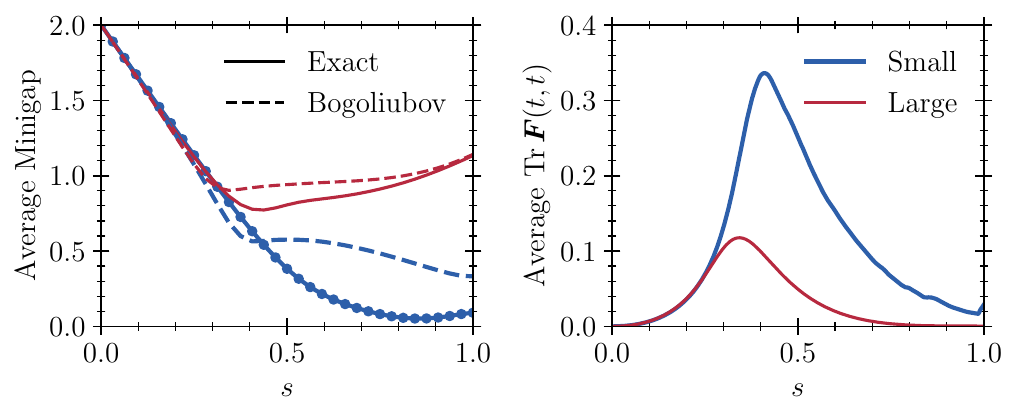}
		\vspace{-0.75cm}\caption
		{\label{fig:gap_flucs_N_11} (Left) Ensemble-averaged minigaps for the two data sets at $N=10$. (Right) Ensemble-averaged sum over the statistical fluctuations for the same parameters as in Fig.~\ref{fig:gap_flucs_N_9}.} 
	\end{center}
\end{figure}

\begin{figure}[htb!]
	\begin{center}
		\includegraphics[width=\linewidth]{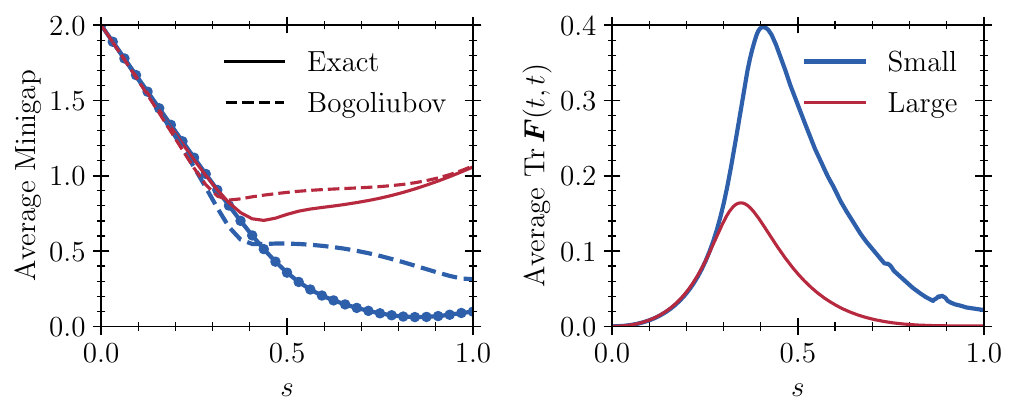}
		\vspace{-0.75cm}\caption
		{\label{fig:gap_flucs_N_13} (Left) Ensemble-averaged minigaps for the two data sets at $N=12$. (Right) Ensemble-averaged sum over the statistical fluctuations for the same parameters as in Fig.~\ref{fig:gap_flucs_N_9}.} 
	\end{center}
\end{figure}

\begin{figure}[htb!]
	\begin{center}
		\includegraphics[width=\linewidth]{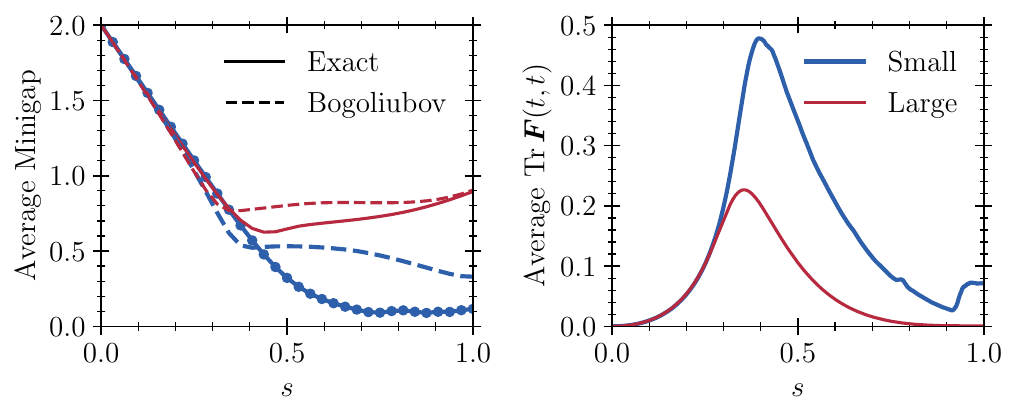}
		\vspace{-0.75cm}\caption
		{\label{fig:gap_flucs_N_15} (Left) Ensemble-averaged minigaps for the two data sets at $N=14$. (Right) Ensemble-averaged sum over the statistical fluctuations for the same parameters as in Fig.~\ref{fig:gap_flucs_N_9}.} 
	\end{center}
\end{figure}

\begin{figure}[htb!]
	\begin{center}
		\includegraphics[width=\linewidth]{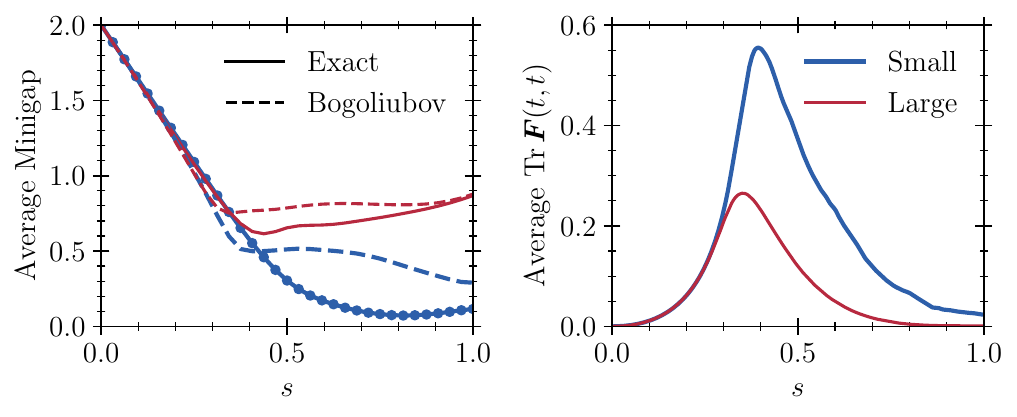}
		\vspace{-0.75cm}\caption
		{\label{fig:gap_flucs_N_17} (Left) Ensemble-averaged minigaps for the two data sets at $N=16$. (Right) Ensemble-averaged sum over the statistical fluctuations for the same parameters as in Fig.~\ref{fig:gap_flucs_N_9}.} 
	\end{center}
\end{figure}

\begin{figure}[htb!]
	\begin{center}
		\includegraphics[width=\linewidth]{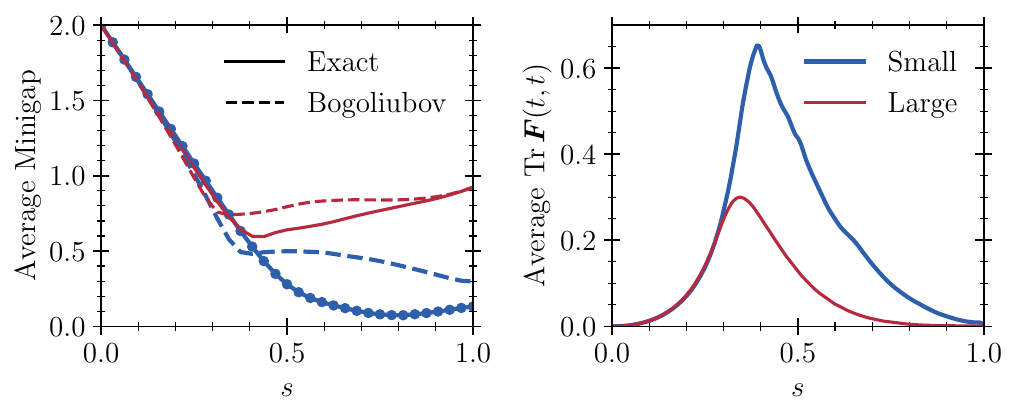}
		\vspace{-0.75cm}\caption
		{\label{fig:gap_flucs_N_19} (Left) Ensemble-averaged minigaps for the two data sets at $N=18$. (Right) Ensemble-averaged sum over the statistical fluctuations for the same parameters as in Fig.~\ref{fig:gap_flucs_N_9}.} 
	\end{center}
\end{figure}

\begin{figure}[htb!]
	\begin{center}
		\includegraphics[width=\linewidth]{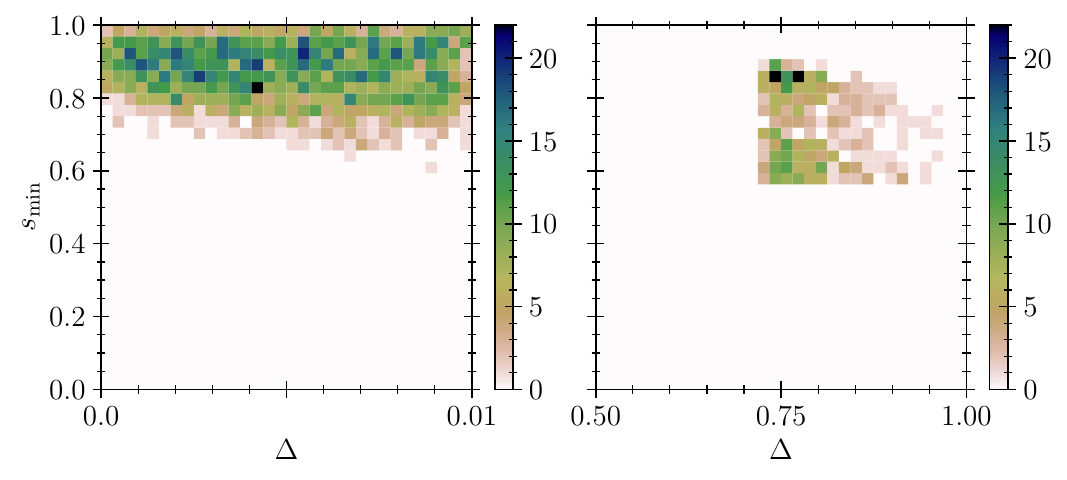}
		\vspace{-0.75cm}\caption
		{\label{fig:gap_loc_size_N_9} minigap size $\Delta$ vs.\ location $s_{\rm min}$ at $N=8$. The cut-off $\Delta_{\rm large} \sim 0.75$ is visible on the right.} 
	\end{center}
\end{figure}

\begin{figure}[htb!]
	\begin{center}
		\includegraphics[width=\linewidth]{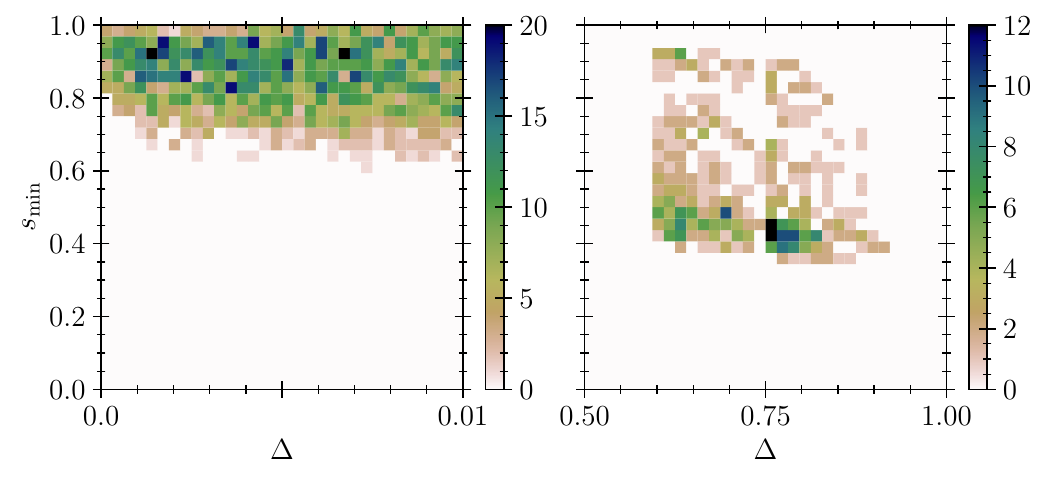}
		\vspace{-0.75cm}\caption
		{\label{fig:gap_loc_size_N_11} minigap size $\Delta$ vs.\ location $s_{\rm min}$ at $N=10$. The cut-off $\Delta_{\rm large} \sim 0.6$ is visible on the right.} 
	\end{center}
\end{figure}

\begin{figure}[htb!]
	\begin{center}
		\includegraphics[width=\linewidth]{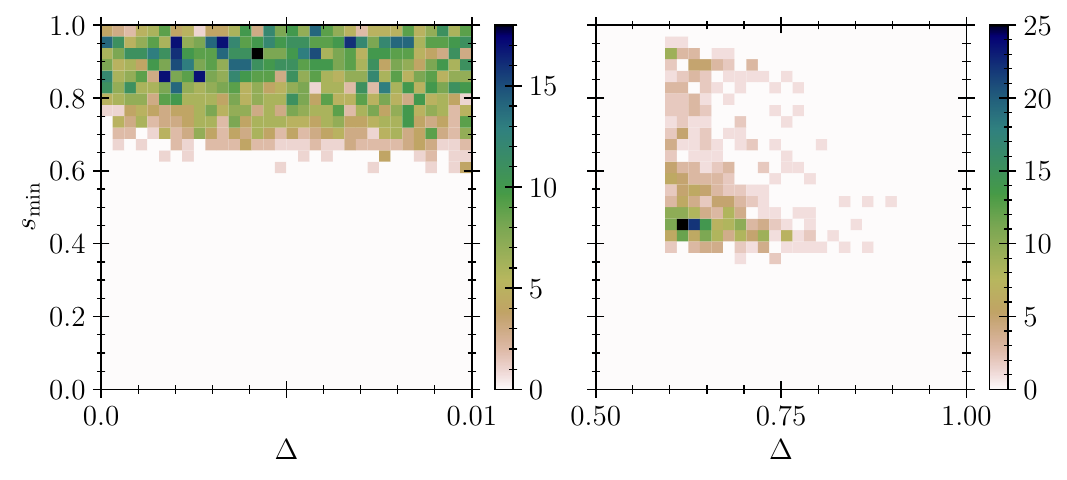}
		\vspace{-0.75cm}\caption
		{\label{fig:gap_loc_size_N_13} minigap size $\Delta$ vs.\ location $s_{\rm min}$ at $N=12$. The cut-off $\Delta_{\rm large} \sim 0.6$ is visible on the right.} 
	\end{center}
\end{figure}

\begin{figure}[htb!]
	\begin{center}
		\includegraphics[width=\linewidth]{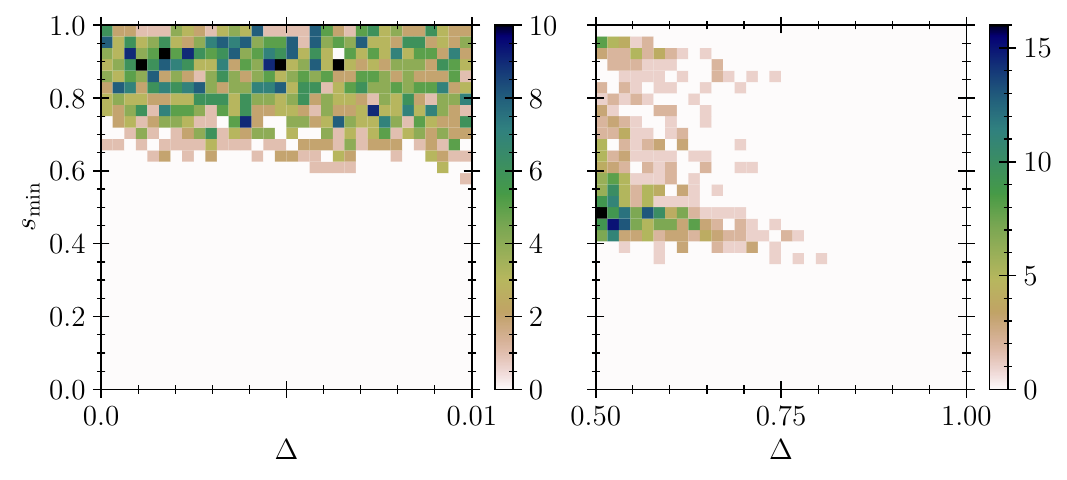}
		\vspace{-0.75cm}\caption
		{\label{fig:gap_loc_size_N_15} minigap size $\Delta$ vs.\ location $s_{\rm min}$ at $N=14$. The cut-off $\Delta_{\rm large} \sim 0.5$ is visible on the right.} 
	\end{center}
\end{figure}

\begin{figure}[htb!]
	\begin{center}
		\includegraphics[width=\linewidth]{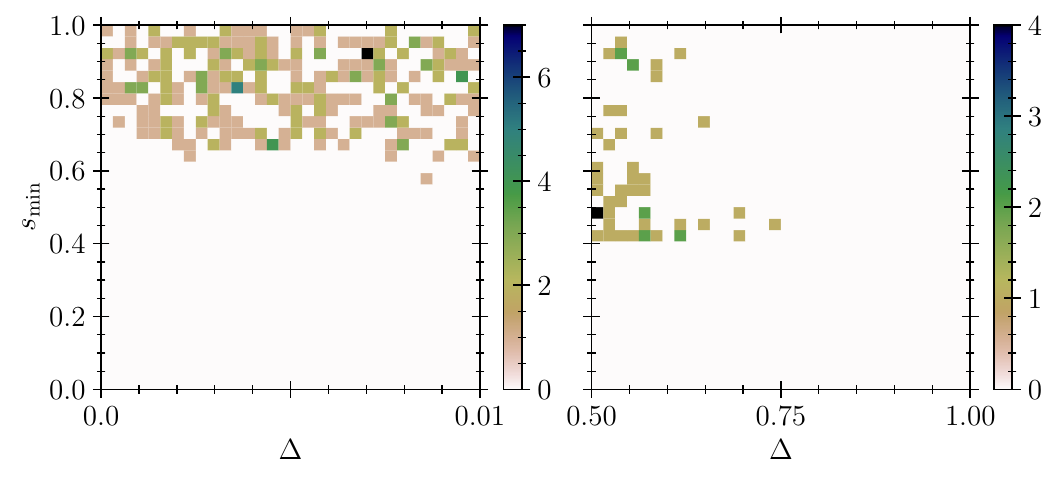}
		\vspace{-0.75cm}\caption
		{\label{fig:gap_loc_size_N_17} minigap size $\Delta$ vs.\ location $s_{\rm min}$ at $N=16$. The cut-off $\Delta_{\rm large} \sim 0.5$ is visible on the right.} 
	\end{center}
\end{figure}

\begin{figure}[htb!]
	\begin{center}
		\includegraphics[width=\linewidth]{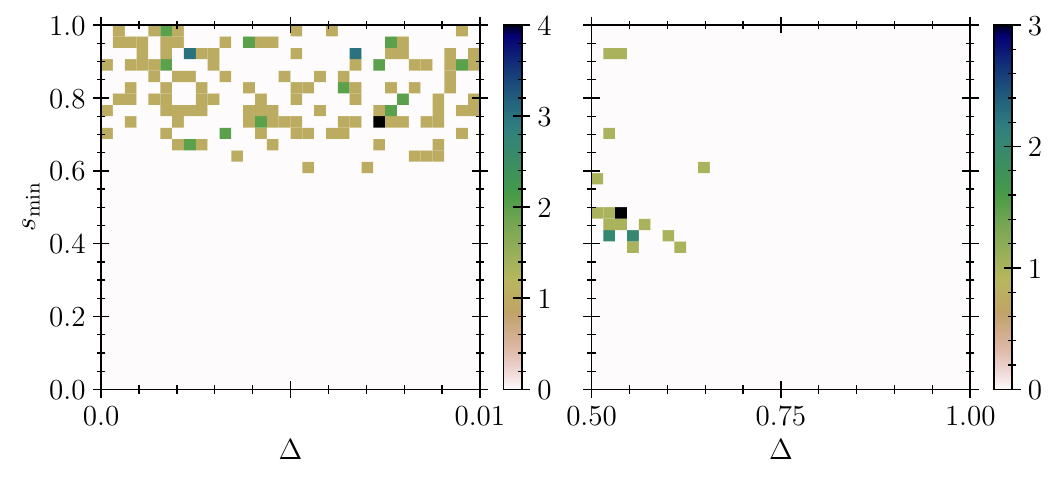}
		\vspace{-0.75cm}\caption
		{\label{fig:gap_loc_size_N_19} minigap size $\Delta$ vs.\ location $s_{\rm min}$ at $N=18$. The cut-off $\Delta_{\rm large} \sim 0.5$ is visible on the right.} 
	\end{center}
\end{figure}

Below, we present plots contrasting the two data sets by means of the ensemble-averaged exact minigaps (obtained from \texttt{ARPACK}'s largest-magnitude eigenvalues~\footnote{Note that, technically, we discard those few instances for which the chosen number of largest-magnitude eigenvalues \textit{does not} contain the ground and first excited states. Hence, the percentage of hard instances for large $N$ is slightly higher than quoted.}), the instantaneous Bogoliubov spectra and the statistical fluctuations. Overall, the magnitude of the fluctuations is found to correlate well with the size of the minigap. The minigap \textit{location}, however, is not clearly deducible from the aggregate fluctuations $\operatorname{Tr} \boldsymbol{F}$. Even so, the fact that the magnitude of the statistical fluctuations correlates so strongly with the size of the minigap can be considered a side result as it confirms the usefulness of the semi-classical approximation to judge the quality of the mean-field solutions. If the bitstring returned by the mean-field is accompanied by small fluctuations, one can be fairly certain that it represents a good solution. This is corroborated by our finding that the adiabatic mean-field dynamics \textit{always} returns the true ground state when applied to the large-minigap subset of our data. Looking to the left panels of Figs.~\ref{fig:gap_flucs_N_9} to \ref{fig:gap_flucs_N_19}, this can also be understood to result from the fact that the true minigap is virtually identical to the approximate one given by the Bogoliubov spectrum. Note the heights of the fluctuation peaks increase as we go from Fig.~\ref{fig:gap_flucs_N_9} to \ref{fig:gap_flucs_N_19}. This also holds for those corresponding to the large-minigap data; however, this is related to the different cut-offs $\Delta_{\mathrm{large}}$ indicated in the right panels of Figs.~\ref{fig:gap_loc_size_N_9} to \ref{fig:gap_loc_size_N_19}. In summary, we believe that the mean-field dynamics in combination with the semi-classical fluctuations could be thought of as a ``first-order diagnostic'' tool that can be used in practice to assess the difficulty of a given problem instance.

\begin{figure*}[htb!]
	\begin{center}
		\includegraphics[width=0.55\textwidth]{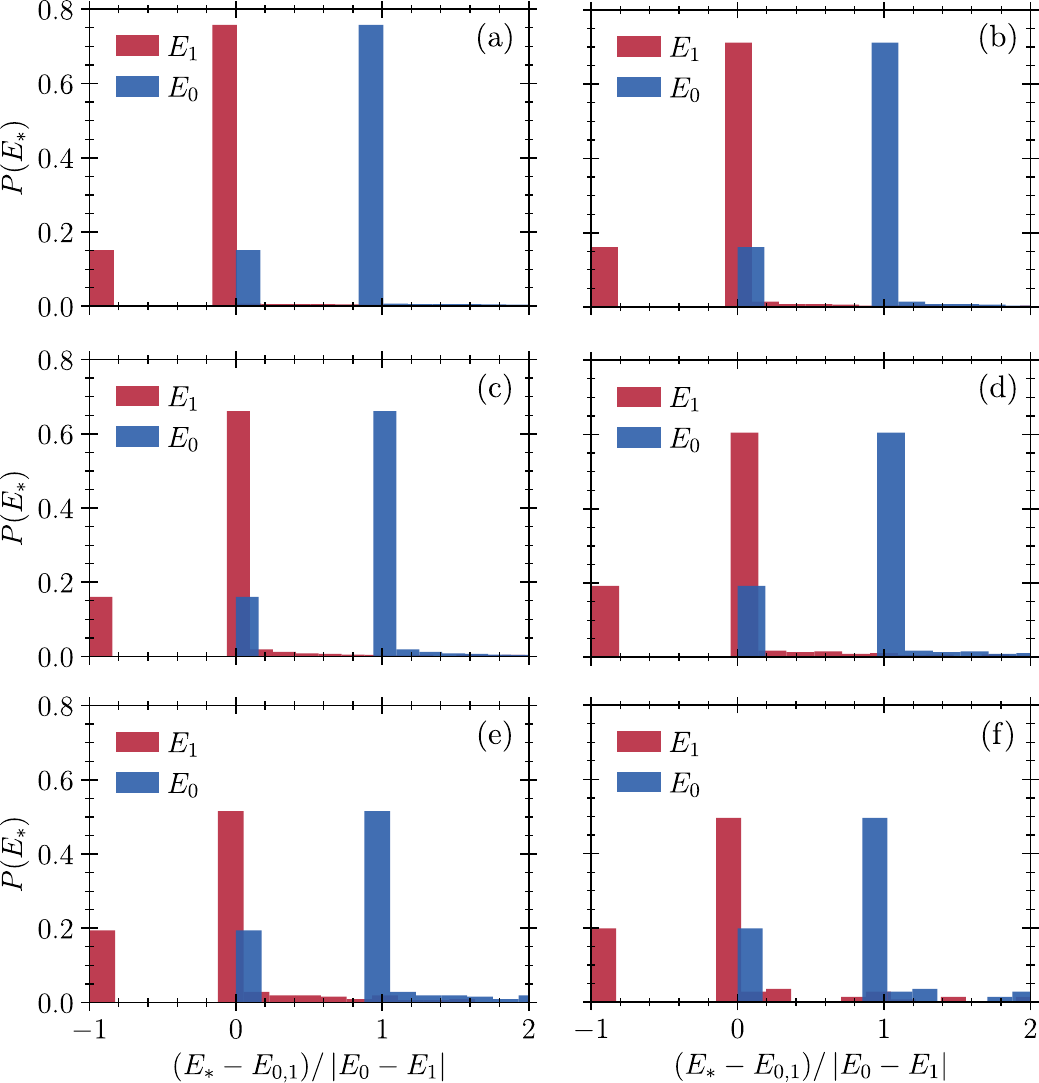}
		\caption
		{\label{fig:mf_sol_hist_all} The energy $E_*$ of the mean-field solution in comparison to the ground- and excited-state energies $E_{0,\,1}$ for the small-minigap instances from (a) $N=8$ to (f) $N=18$. The exact solution (i.e.\ $E_* - E_0 \approx 0$) is recovered for $\{384, 412, 368, 247, 60, 27\}$ instances, respectively, with the total ensemble sizes given in Tab.~\ref{tab:SK_small}. Contrary to this, for the large-minigap instances, the exact solution is recovered in \textit{all} cases.} 
	\end{center}
\end{figure*}

The present data also furnish a more detailed understanding of why the mean-field approximate optimization algorithm~\cite{Misra_2023} scales so well for the SK model. While the fraction of instances for which the true solution is recovered seems to increase systematically (s.~Fig.~\ref{fig:mf_sol_hist_all}), the scatter to higher excited states also goes up. Now it is known that, as $N\to\infty$, the spectral weight of the SK model accumulates close to $E_0(T_f)$~\cite{Mueller_2012}. In this way, even though the mean-field algorithm converges to excited states, the absolute distance in energy may still go down asymptotically, simply because more and more almost degenerate states appear close to the true ground state.

\bibliography{refs}

\end{document}